# Tunable contact angle hysteresis on micropatterned surfaces

Damien Debuisson, Vincent Senez and Steve Arscott,[a]

*Institut d'Electronique, de Microélectronique et de Nanotechnologie (IEMN), University of Lille, CNRS UMR-8520, Cite Scientifique, Villeneuve d'Ascq, France.*

Micropatterned surfaces composed of concentric circular defects having a smooth trench-like profile are formed using a photoresist (SU-8). When an evaporating droplet encounters the micropatterned surface an evaporation phase is observed consisting of distinct discontinuities and steps in the droplet wetting contact angle and base radius respectively. The addition of gaps into the circular defects enables tuning of the contact angle hysteresis; the receding contact angle of fluorocarbon coated SU-8 can be tuned between 34.6° and 89.1° and that of SU-8 surfaces from 5.6° to 43.3° depending on the gap length. In addition, a model is developed which accurately predicts the observed behavior.

[a]Electronic mail: steve.arscott@iemn.univ-lille1.fr.

Wetting is of great scientific interest.[1-6] In particular contact angle hysteresis,[7-11] which is the difference between the advancing contact angle and the receding contact angle of a moving contact line. Contact angle hysteresis exists because surfaces are not homogenous; models based on roughness[1,9] and zones of differing surface energies[2,10] can be used to interpret experimental data. Control of contact angle hysteresis has importance in several modern micro/nanotechnology applications[12-15].

Consider Fig. 1(a) which shows a droplet pinned to a circular "defect"[3,10] having a radius $r_d$ which contains a gap having a lateral length ($x$ direction) $G$. Figs. 1(b) and 1(c) show side-views of the droplet cut along $BB'$ and $CC'$. $\theta_{rG}^*$ is the contact angle when the droplet depins from the circular defect at points $p$ and $p'$; we call this the effective receding contact angle for a given $G$; this is the *minimum* contact angle a droplet can have on such defects. $\theta_r$ is the receding contact angle of the droplet on a surface without a circular defect. As we will see, $\theta_{rG}^*$ can vary between $\theta_r$ ($G = 2\pi r_d$) and $\theta_r^*$ ($G = 0$)[16]. Considering the small meniscus formed in the locality of the gap, we have two radii of curvature $r_2$ and $r_3$, with $r_1$ being the radius of curvature of the droplet, cf. Fig. 1(b). Using the Young-Dupré equation[17,18] we have the net force in the $y$-direction:

$$F_y = \gamma \int_\Gamma dl \mathbf{e}_y \cdot \mathbf{n} \cos\overline{\theta} \quad (1)$$

where, $\gamma$ is the liquid-vapor surface energy, $\overline{\theta}$ is the local contact angle, $\mathbf{n}$ is the unit vector normal to the contact line in the $x$-$y$ plane, $\mathbf{e}_y$ is the unit vector pointing in the $y$-direction, $l$ is the arc length parameter of the contact line, $\Gamma$ is the contact line which is composed of: (i) a defect pinned portion $\Gamma_d$ and (ii) a gap associated portion $\Gamma_G$. When the droplet depins: $\overline{\theta} = \theta_{rG}^*$ along $\Gamma_d$ and $\overline{\theta} = \theta_r$ along $\Gamma_G$. By solving Equation 1 over the contact line $\Gamma$ we have:

$$F_y = \gamma G(\cos\theta_{rG}^* - \cos\theta_r) \quad (2)$$

by assuming that the depinning of the liquid is independent of $G$ we have a constant force $f_d$ associated with the defect[10]. Thus, the effective receding contact angle (for a given $G$) can be written:

$$\theta_{rG}^* = \cos^{-1}\left(\frac{f_d}{\gamma G} + \cos\theta_r\right) \quad (3)$$

Note that Equation 2 is a solution for an arbitrary shaped defect. We can now determine expressions for the radii of curvature of the meniscus. Near to the gap the liquid forms a small meniscus of radius $r$ (in the $x$-$y$ plane). This radius is the intersection of the sphere of radius $r_3$ and the $x$-$y$ plane. Note that $r_1$ and $r_2$ are positive whilst $r$ and $r_3$ are negative. The Young-Laplace equation[18,19] allows us to write:

$$\frac{2}{r_1} = \frac{1}{r_2} + \frac{1}{r_3} \quad (4)$$

Thus, in terms of the parameters: $r$, $r_d$, $\theta_{rG}^*$ and $\theta_r$, simple geometry can be employed to give the following:

$$r_1 = \frac{r_d}{\sin\theta_{rG}^*} \quad (5)$$

$$r_2 = \left(\frac{2\sin\theta_{rG}^*}{r_d} - \frac{\sin\theta_r}{r}\right)^{-1} \quad (6)$$

$$r_3 = \frac{r}{\sin\theta_r} \quad (7)$$

In order to test the above reasoning we have fabricated micropatterned smooth defects using lithographic methods. The photoresist SU-8 (Microchem, USA) was used to form circular trench-like defects; SU-8 is very versatile for microfluidics[20,21]. The micropatterned surfaces were fabricated on 3-inch diameter silicon wafers (Siltronix, France) using SU-8 2010. The photoresist was spin coated (3300 rpm/1000 rpm s$^{-1}$/30 s) and prebaked (95°C for 4 minutes) which resulted in a thickness of 9.56 µm (±0.33 µm). The micropatterns were defined using a single photomask; optimized lithography[22] (7.7 s/10 mW cm$^{-2}$) using a 1 µm linewidth caused Fresnel diffraction underneath the mask[23] which results in the smooth trench-like defects as show in Fig. 2. Once the SU-8 surfaces had been prepared a fluorocarbon layer (FC) (thickness ~100 nm) could be deposited using a C$_4$F$_8$ plasma (STS Ltd., UK)

If there are no gaps in the circles the receding contact angle of such a micropattern has an *effective* receding contact angle $\theta_r^*$ which depends on the *profile* of the trench-like defect[16]. It will be seen here that by adding gaps the value of $\theta_r^*$ can be effectively tuned. The concentric circles had a spacing $D$ of 50 and 100 µm and gap length $G$ which ranged from 20 to 200 µm. In each pattern, the largest concentric circle had a diameter of 2 mm; the capillary length[18] $\sqrt{\gamma/\rho g}$ of water ($\gamma$ = 72.8 mJ m$^{-2}$ and $\rho$ = 998.2 kg m$^{-3}$)[24] is around 2.7 mm so gravity effects can be ignored in the interpretation of the results. In addition to testing the micropatterns with the gaps we measured the receding contact angle of water on the SU-8 surface used for the study (43.3°±1.6°) and the SU-8 coated with FC (89.1°±2.5°) and on a micropattern without gaps, i.e. closed concentric circles SU-8 (5.6°±1.3°) and SU-8 coated with FC (34.6°±3.4°).

The contact angle measurements were performed using a Drop Shape Analysis System DSA100 (Kruss, Germany). The experiments were performed in a class ISO 5/7 cleanroom ($T$ = 20°C±0.5°C; $RH$ = 45%±2%). A single drop (*vol* ~2 µL) of deionized water was placed over a micropattern using a pipette and allowed to evaporate (unforced). The evolution of the droplet shape from above, Fig 1(a), and plan view (see the *BB'* cross-section in Fig. 1(b).) was recorded as a function of time as the droplet evaporated. At least 10 measurements were conducted for each point and error bars are the standard deviation. Analysis of the images allowed us to extract the value of $r$ at the moment of droplet depinning from the circular pattern initially at points *p* and *p'* (cf. Fig. 1(a))

Fig. 3 shows optical microscope images of the evolution of the water droplet shape (plan view) as a function of time (~80 s). The micropattern here is concentric circles spaced at a distance equal to 50 µm and a gap equal to 100 µm. The surface is SU-8 coated with 100 nm of FC. The recorded times between the images are 20 s for (i) to (iii) and 60 s between (iv) to (v). In contrast, the droplet depinning, see the transitions from (i) to (ii), (iii) to (iv) and (v) to (vi), occurs over a relatively short period in < 200 ms. Figs. 3(b) and 3(c) show the evolution of the droplet profile (side-view) as a function of time for two different values of $G$. As the droplet evaporates it becomes pinned on a circular defect with a meniscus forming around the gap. As the contact line moves down the trench-like defect, the effective contact angle diminishes at constant base radius[16]. At the moment of depinning, the droplet depins from the gap (points *p* and *p'* in Fig 1(a)) and evaporates further to pin onto a smaller concentric circle, the process repeating itself until full evaporation of the water.

Fig. 4 shows plots of the droplet contact angle and base radius as a function of time. The two well-known evaporation phases[25] are clearly visible (indicated on Fig. 4 as *A* and *B*). The first phase *A* corresponds to evaporation at constant droplet base radius with reducing contact angle whilst the second phase *B* to corresponds to reducing droplet base radius at constant contact angle. These two phases are observed until the contact line portion $\Gamma_d$ is entirely pinned to a single circle defect. Once the droplet is pinned a new phase *C* is observed where the contact angle and the base radius of the droplet evolve in distinct jumps and steps respectively as the droplet pins to and depins from successive circular defects. Note that the phase *B'* (see above[25]) indicated in Fig. 4 corresponding to the initial droplet depinning from a circle ($r_d$ = 0.5 mm) finishes *between* two circular defects as the droplet depins at a larger contact angle than $\theta_{rG*}$ due to a surface impurity. In a first approximation by assuming that the jumps in the contact angle occur at constant volume we can write the following expression relating the droplet base radius and the droplet contact angle:

$$\left(\frac{r_d^*}{r_p}\right)^3 = \frac{\tan\frac{\theta_p}{2}\left(3+\tan^2\frac{\theta_p}{2}\right)}{\tan\frac{\theta_{rG}^*}{2}\left(3+\tan^2\frac{\theta_{rG}^*}{2}\right)} \quad (8)$$

where $r_d^*$ is the droplet base radius at depinning, $r_p$ is the new radius, $\theta_{rG}^*$ is described above and $\theta_p$ is the new contact angle.

Fig. 5(a) shows plots of $\theta_{rG}^*$ as a function of $G$ for the two different surfaces measured (SU-8 and SU-8 coated with FC). Equation 3 is also plotted on Fig. 5(a) using the value of $f_d$ extracted from the measurements and the value of $\theta_r$ measured on flat surfaces. The value of $f_d$ was calculated from the data to be 534.4 nN±9.5 nN and 283.7 nN±14.9 nN for SU-8 coated with FC and SU-8 respectively. As $G$ increases then the value of $\theta_{rG}^*$ tends to $\theta_r$. However, as the value of $G$ diminishes clearly our model does not predict the value of $\theta_r^*$ for a circle having $G$ = 0; 5.6°±1.3° in the case of SU-8 and 34.6°±3.4° in the case of SU-8 coated with FC[16]. The physical reason for this is that the value of the $\theta_{rG}^*$ cannot be smaller that the value measured on a closed concentric circle ($G$ = 0). Thus in reality, $\theta_{rG}^*$ will follow the horizontal dashed lines (red and blue) indicated in Fig. 5(a). In this way we are able to predict the minimum value of $G$ which can have an influence on the value of $\theta_{rG}^*$; this value is 9 µm for SU-8 coated with FC and 14.5 µm for SU-8, for the defect profile[16] studied here. A plot of $\cos\theta_{rG}^*$ - $\cos\theta_r$ versus $1/\gamma G$ shown in Fig. 5(b) reveals a straight line passing through the origin which is predicted by our model; a first order polynomial fit to the data yields a defect force $f_d$ of 548 nN and 257 nN on SU-8 coated with FC and SU-8 respectively. By measuring the value of the meniscus radius $r$ (cf. Fig. 1(c)) at the moment of depinning and knowing the value of $\theta_{rG}^*$ one can calculate the values of the radii of curvature in Fig. 1(c) at each value of $G$ on a given surface (for $r_d$ = 0.5 mm). Tables I and II presents numerical results of the study.

[1]R. N. Wenzel Ind. Eng. Chem. **28**, 988 (1936).
[2]A. B. D. Cassie and S. Baxter, Trans. Faraday Soc. **40**, 546 (1944).
[3]P. G. De Gennes, Rev. Mod. Phys. **57**, 827 (1985).
[4]R. D. Deegan, O.Bakajin, T. F. Dupont, G. Huber, S. R. Nagel, and T. A. Witten, Nature **389**, 827 (1997).
[5]L. Feng *et al.* Adv. Mater. **14**, 1857 (2002).
[6]A. Lafuma, and D. Quéré, Nature Materials **2**, 457 (2003).
[7]F.E. Bartell and J.W. Shepard, J. Phys. Chem. **57**, 455 (1953).
[8]R. E. Johnson and R. H. Dettre, J. Phys. Chem. **68**, 1744 (1964).
[9]C. Huh and S.G. Mason, J. Colloid and Interface Sci. **60** 11 (1977).
[10]J. F. Joanny and P. G. De Gennes, J. Chem. Phys. **81**, 552 (1984).
[11]Y. L. Chen, C. A. Helm and J. N. Israelachvili, J. Phys. Chem. **95**, 10736 (1991).
[12]R. Blossey, Nature Mater. **2**, 301 (2003).
[13]D. J. Harrison, K. Fluri, K. Seiler, Z. Fan, C. S. Effenhauser and A. Manz, Science **261**, 895 (1993).
[14]G. M. Whitesides and B. Grzybowski, Science **295**, 2418 (2002).
[15]J. B. Hannon, S. Kodambaka, F. M. Ross and R. M. Tromp, Nature **440**, 69 (2006).
[16]D. Debuisson, R. Dufour, V. Senez and S. Arscott, submitted (online at arxiv.org/abs/1101.0915v1)
[17]T. Young, Philos. Trans. Soc. London, **95**, 65 (1805).
[18]*Capillarity and Wetting Phenomena: Drops, Bubbles, Pearls, Waves*, P.G. de Gennes, F. Brochard-Wyart and D. Quéré, Springer Science, USA (2003).
[19]de Laplace, P-S. Œuvres complètes de Laplace, t. IV, supplément au livre X du traité de la mécanique céleste, p. 394 – 2e supplément au livre X, p. 419, Ch. I. (in French).

[20] J. Carlier, S. Arscott, V.Thomy, J. C. Fourier, F.Caron, J. C. Camart, C. Druon, and P. Tarbourier, J. Micromech. Microeng. **14**, 619 (2004).
[21] S. Arscott, S. Le Gac, C. Druon, P. Tabourier and C. Rolando, J. Micromech. Microeng. **14**, 310 (2004).
[22] M. Gaudet, J. C. Camart, L. Buchaillot and S. Arscott, Appl. Phys. Lett. **88**, 024107 (2006).
[23] Y Cheng, C-Y Lin, D-H Wei, B. Loechel and G. Gruetzner, J. MEMS **8**, 18 (1999).
[24] *CRC Handbook of Chemistry and Physics*, 69th Edition, Edited by R. C. Weast, CRC Press, Florida, USA.
[25] R. G. Picknett and R. Bexon, J. Colloid Interface Sci. **61**, 336 (1977).

FIG. 1. (a) Top view of a droplet pinned to a circular shaped defect of radius $r_d$ having a gap of lateral length $G$. (b) cross-section of the same droplet along the $BB'$ plane as indicated in Fig. 1(a). (c) cross-section of the meniscus formed by the gap along the $CC'$ plane as indicated in Fig. 1(a).

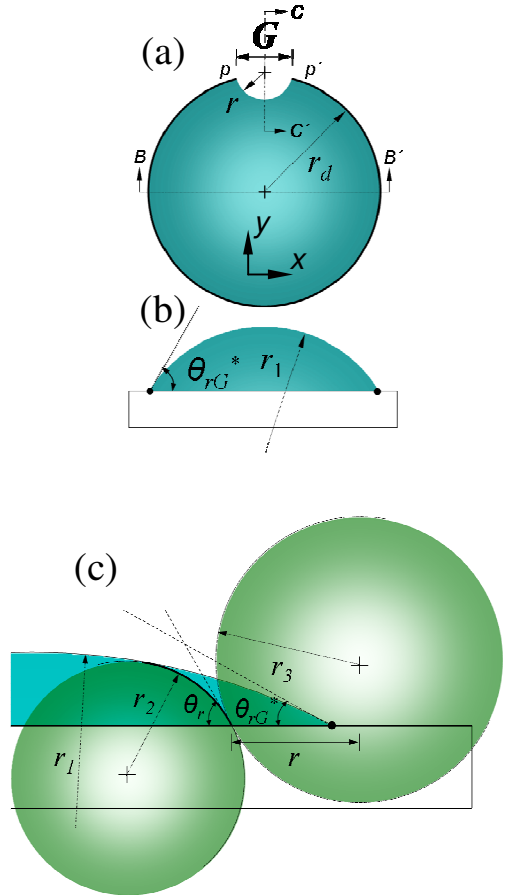

FIG. 2. Scanning electron microscopy image of the cross-section of a smooth trench-like defect fabricated using photolithography of the photoresist SU-8.

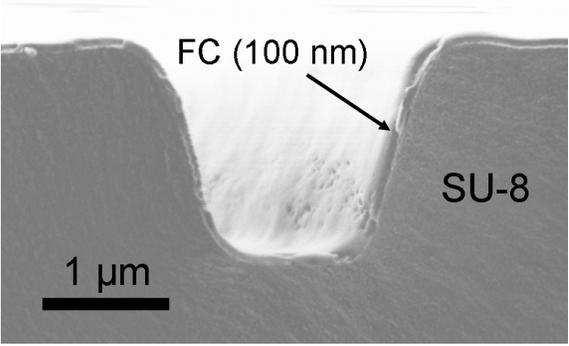

FIG. 3. Evolution of the droplet shape as a function of time (a) plan view for H$_2$O droplet on SU-8 coated with FC surface containing smooth trench like circular defects shown in Fig. 2. (Circle spacing $D = 50$ µm, gap length $G = 100$ µm). Effect of gap length on the contact angle hysteresis for (b) $D = 100$ µm, $G = 0$ and (c) $D = 100$ µm, $G = 50$ µm. (b) and (c) are an SU-8 surface smooth trench like circular defects.

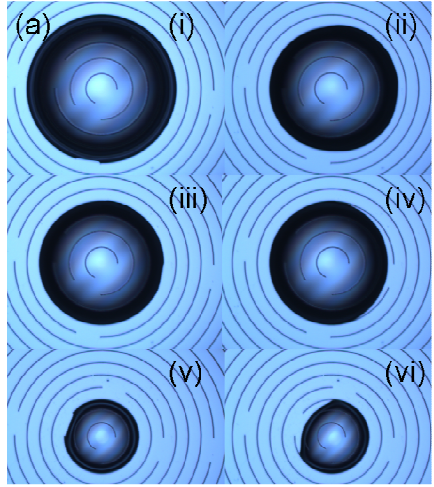

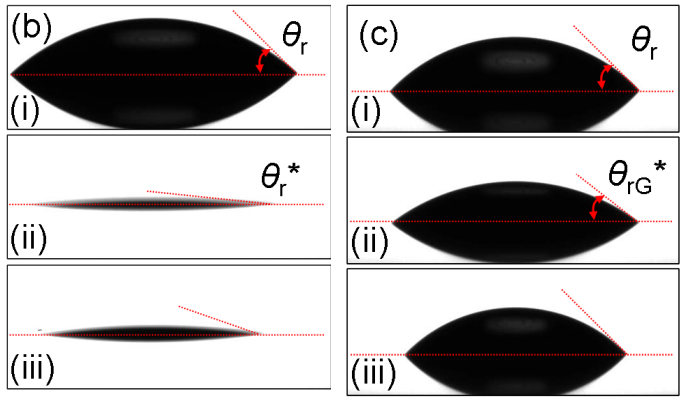

FIG. 4. Variation of droplet contact angle and base radius as a function of time for an evaporating droplet (H$_2$O) on a micropatterned surface (SU-8 coated with FC with $D$ = 50 µm and $G$ = 20 µm).

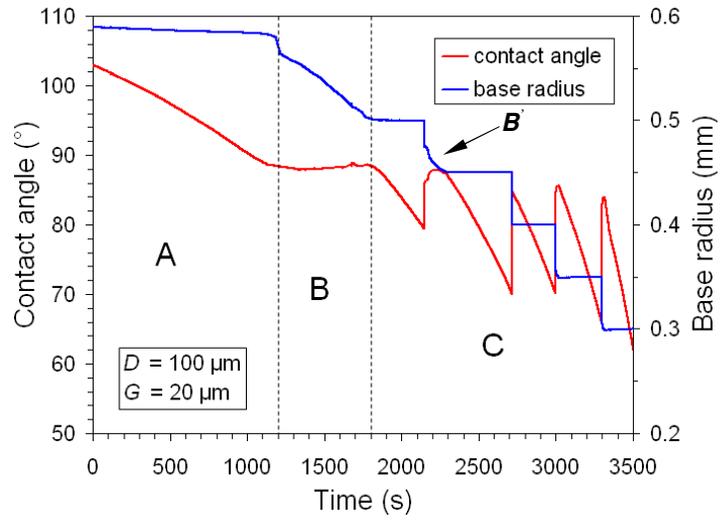

FIG. 5. (a) Variation of the effective receding contact angle $\theta_{rG}^*$ of the droplet (red squares = $H_2O$ on SU-8 coated with FC, blue triangles = $H_2O$ on SU-8) as a function of gap length $G$ for micropatterned surfaces containing smooth trench-like defects shown in Fig. 2. The dashed black line corresponds to the model (Equation 3) by fitting the average defect force $f_d$ calculated from the measurements. (b) A plot of $\cos\theta_{rG}^* - \cos\theta_r$ versus the reciprocal of surface tension-gap length product $\gamma G$. The dashed line represents a first order polynomial fit.

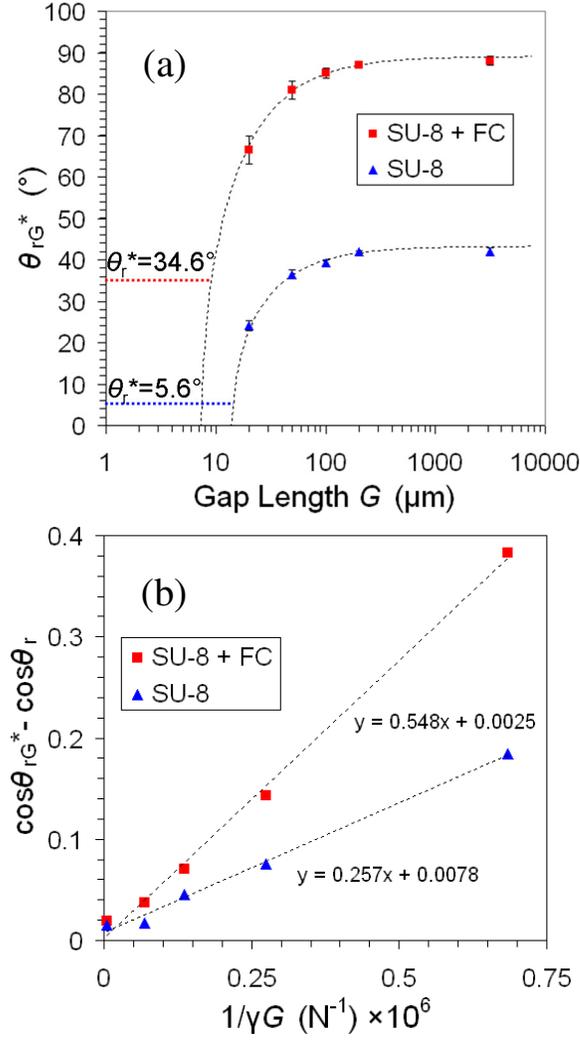

| SU-8 | Experimental | | | Calculated | |
|---|---|---|---|---|---|
| $G$ | $\theta_{rG}$*(°) | $r_1$ (μm) | $r$ (μm) | $r_2$ (μm) | $r_3$ (μm) |
| 20 | 24.2(±1.1) | 1219.7 | -16.8 | 23.6 | -24.5 |
| 50 | 36.5(±0.8) | 840.6 | -75.7 | 87.4 | -110.4 |
| 100 | 39.3(±0.4) | 789.4 | -288.7 | 203.7 | -421.0 |
| 200 | 41.8(±0.9) | 750.2 | -291.0 | 199.1 | -424.3 |

Table I: Summary of results for SU-8.

| SU-8+FC | Experimental | | | Calculated | |
|---|---|---|---|---|---|
| $G$ | $\theta_{rG}$* (°) | $r_1$ (μm) | $r$ (μm) | $r_2$ (μm) | $r_3$ (μm) |
| 20 | 66.5(±2.1) | 66.5 | 545.2 | -14.9 | 14.1 |
| 50 | 80.9(±1.1) | 80.9 | 506.4 | -87.3 | 64.9 |
| 100 | 85.1(±0.4) | 85.1 | 501.8 | -553.7 | 172.7 |
| 200 | 87.0(±1) | 87.0 | 500.7 | 1366.0 | 306.5 |

Table II: Summary of results for SU-8 coated with FC